# Magnetostrictive Fe$_{73}$Ga$_{27}$ nanocontacts for low-field conductance switching

**U. M. Kannan[1], S. Kuntz[2], O. Berg[2], Wolfram Kittler[2], Himalay Basumatary[3], J. Arout Chelvane[3], C. Sürgers[2] and S. Narayana Jammalamadaka[1]\***

[1]Magnetic Materials and Device Physics Laboratory, Department of Physics, Indian Institute of Technology Hyderabad, Hyderabad 502 285, India

[2]Physikalisches Institut, Karlsruhe Institute of Technology, Wolfgang Gaede Str. 1, Karlsruhe, 76131, Germany

[3]Defence Metallurgical Research Laboratory, Hyderabad 500058, India

Corresponding authors: surya@iith.ac.in

**Abstract:**

The electrical conductance $G$ of magnetostrictive nanocontacts made from Galfenol (Fe$_{73}$Ga$_{27}$) can be reproducibly switched between 'on' and 'off' states in a low magnetic field of ~ 20 – 30 mT at 10 K. The switching behavior is in agreement with the magnetic field dependence of the magnetostriction inferred from the magnetization behavior, causing a positive magnetostrictive strain along the magnetic field. The repeated magnetic-field cycling leads to a stable contact geometry and to a robust contact configuration with a very low hysteresis of ~ 1 mT between opening and closing the contact due to a training effect. Non-integral multiples of the conductance quantum $G_0$ observed for $G > G_0$ are attributed to electron backscattering at defect sites in the electrodes near the contact interface. When the contact is closed either mechanically or by magnetic field, the conductance shows an exponential behavior below $G_0$ due to electron tunneling. This allows to estimate the magnetostriction $\lambda = 4 \times 10^{-5}$ at 10 K. The results demonstrate that such magnetostrictive devices are suitable for the remote control of the conductance by low magnetic fields in future nanotechnology applications.

## Introduction

Research is driven by the pursuit of developing smaller, faster, cheaper, and more capable electronic devices as silicon-based technologies are reaching their limits. In this pursuit, electronic transport through single atoms or molecules has gained much attention during the past years[1]. Experimentally, electronic quantum transport has been investigated by techniques like scanning tunneling microscopy (STM) [2], table top set-ups [3], and mechanically controlled break junctions (MCBJ) [4-6]. STM and MCBJ allow controlling the relative displacement of two electrodes with a resolution of a few picometer by using a piezoelectric element. In a MCBJ, a thin wire or micro structured contact is mechanically broken thus forming two metal contacts which are separated by a narrow gap. In the ballistic regime, where the conductance $G$ is only a few conductance quanta $G_0 = 2e^2/h$ (e: electron charge, h: Planck's constant), the conductance is usually described by the Landauer-Büttiker theory[7].

Besides the large number of experiments performed on nonmagnetic metals, ferromagnetic MCBJs and nanowires have been explored by measurements of magnetoresistance and conductance quantization [8-11]. Tuning the magnetoresistance of ferromagnetic nanocontacts by utilizing the magnetostriction has been reported earlier [12]. Furthermore, magnetostriction has been used to switch the conductance of $Dy$[13] and $Tb_{0.3}Dy_{0.7}Fe_{1.95}$ (Terfenol-D) [14] MCBJ in magnetic fields above 100 mT at low temperature and room temperature, respectively.

The magnetostrictive strain $\lambda$ - the relative length change $\Delta l/l$ in the direction of the magnetization - is associated with the magnetization process and can be changed by application of a magnetic field $H$ [15,16,17]. Magnetostriction is particularly strong in magnetic transition-metals and rare-earth elements. A considerable enhancement of $\lambda$ compared to 3d transition metals has been observed for $Fe_{1-x}Ga_x$ alloys [18,19] for which a large magnetostriction $\lambda_{001} > 3 \times 10^{-4}$ has been reported in the <100> direction at room temperature, depending on $x$ and on the quenching conditions [20]. This corresponds to a tenfold increase of $\lambda$ compared to bulk bcc-Fe ($\lambda_{001} = 3 \times 10^{-5}$, $\lambda_{111} = -1.6 \times 10^{-5}$) [21]. Recently, non-volume conserving or non-Joulian magnetostriction has been reported for $Fe_{1-x}Ga_x$ single crystals [22]. The advantages of $Fe_{1-x}Ga_x$ over the giant magnetostrictive Terfenol-D are its large magnetostriction at low magnetic fields, low magnetic hysteresis, good ductility, and excellent mechanical properties. This makes $Fe_{1-x}Ga_x$ very attractive for sensing and actuating applications in micro-electromechanical systems (MEMS) [23]. Hence, $Fe_{1-x}Ga_x$ could be ideally suited for tuning the conductance of a MCBJ device by a low magnetic field and for controlling the gap between the contacts in a remote way. In this work, we demonstrate the conductance switching of a magnetostrictive $Fe_{73}Ga_{27}$ ribbon at 10 K in a low magnetic field of 30 mT. Furthermore, by controlling the conductance of the MCBJ device mechanically as well as by magnetic field in the regime of electron tunneling we are able to estimate the effect of magnetic field on the electrode separation and the value of $\lambda$.

**Experimental Methods**

A polycrystalline $Fe_{73}Ga_{27}$ ingot was prepared by employing an arc furnace and melting the constituent elements Fe (purity 99.9 wt%) and Ga (purity 99.95 wt%) under argon atmosphere. The melting was performed several times to ensure homogeneity of the alloy. The weight loss after the final melting was less than 0.5 %. Subsequently, ribbons were prepared from the ingot by employing a single roller melt-spinning unit under argon atmosphere. The copper-wheel speed over which the liquid metal quenched was maintained at a constant speed of 34 m/sec. Phase purity of the ribbons was confirmed by X–ray diffraction (XRD) using Cu-K$_\alpha$ radiation. In order to study the microstructure of the Fe$_{73}$Ga$_{27}$ ribbon, scanning electron-microscopy (SEM) images were recorded with various magnifications using a FEI Quanta SEM. Magnetization curves were measured in a vibrating-sample magnetometer (VSM, Oxford Instruments) at 10 K with the magnetic field oriented parallel to the plane of the ribbon. The VSM vibrating frequency of the sample was 55 Hz at 0.2 mm amplitude. Magnetostriction can be measured by the strain-gauge technique [24] or by a modified STM [25]. In the present case, we used commercially available strain gauges (Micro-Measurements Group Inc. U. S. A), made from temperature compensated 120-$\Omega$ Karma foil with negligible magnetoresistance. Cynoacrylate cement (M-bond 200 or Anabond 202) was used to fix the strain gauge to the sample. A copper wire with gauge 38 was used to connect the leads of the strain gauge to a Wheatstone's bridge network as one of the resistors. The magnetostriction was calculated using

$$\lambda = \frac{\Delta l}{l} = \frac{4\Delta E}{VK}$$

where $\Delta E$ is the unbalanced bridge voltage, $V$ is the bridge excitation voltage, and $K$ is a calibration factor. The magnetic field direction was parallel to the plane of the ribbon.

Figure 1 shows a schematic of the MCBJ. The Fe$_{73}$Ga$_{27}$ MCBJ was fabricated by tightly fixing the ends of the ribbon (2 mm x 8 mm) to a flexible copper–bronze substrate with Stycast 2850FT epoxy to minimize mechanical instabilities. Cu wires were attached to the ends of the ribbon for electrical measurements. In order to avoid direct electrical contact between the ribbon and substrate, the substrate was coated by a 2-$\mu$m thick durimide film. The MCBJ device was inserted into a $^4$He bath cryostat and the sample chamber was purged many times with helium gas before breaking the wire. After initial breaking of the ribbon by bending the substrate with a motor-driven pushing rod, the electrode distance $\Delta x$ was further controlled with picometer precision by a piezo stack. At a constant current of 1 µA, the voltage was continuously measured across the junction in order to monitor the $G(\Delta x)$ dependence. A superconducting Helmholtz coil generates a magnetic field along the x direction in the plane of the substrate. All measurements were done at a temperature $T = 10$ K.

**Results and Discussion**

Figure 2(a) shows the XRD pattern for the rapidly quenched $Fe_{73}Ga_{27}$ ribbon confirming the formation of the bcc A2-phase. The dependence of the magnetostrictive strain $\lambda$ on the magnetic field is shown in Fig. 2(b). The magnetostriction is positive and the saturation value $\lambda_S = 3.3$ x $10^{-4}$ reached at $\mu_0 H \sim 0.18$ T is in agreement with values reported earlier [26]. The SEM image Fig. 2(c) clearly shows a distribution of grains and grain boundaries. Beyond that the sample is devoid of noticeable defects. Fig. 2(d) shows the photograph of a MCBJ device fabricated from a $Fe_{73}Ga_{27}$ ribbon.

The conductance switching of a $Fe_{73}Ga_{27}$ nanocontact at 10 K is shown in Fig. 3(a). By increasing the magnetic field from zero in the "open" configuration (zero conductance), the conductance suddenly jumps to a finite value $G = 125$ $G_0$ at $\mu_0 H = 28$ mT representing a closed contact. From the "close" state the contact can be reopened by reducing the field, leading to a jump of the conductance to zero at a field only $\sim 5$ mT lower than for closing the contact. A similar behavior is observed when the field is reversed in the negative field direction. The opening and closing of the contact could be reproducibly repeated several times. The magnetic-field induced switching is attributed to the magnetostrictive strain of the $Fe_{73}Ga_{27}$ causing an extension of both electrodes in increasing field along the field direction, cf. Fig. 2(b). It is important to note that in addition to the small hysteresis the "switching field" of the device at $T = 10$ K is well below 100 mT while opening and closing the contact which demonstrates that MCBJs made from $Fe_{73}Ga_{27}$ can be advantageously used in high sensitivity magnetic-field applications.

The magnetization curve $M(H)$ shown in Fig. 3(b) reveals a soft ferromagnetic behavior with a hysteresis of ~ 15 mT width. In order to correlate the $G(H)$ with the $M(H)$ behavior, we assume the relative magnetostrictive strain $\lambda/\lambda_S$ along the magnetization axis x to be proportional to $(M/M_S)^2$, see Fig. 3(c). $\lambda_S$ and $M_S$ are the magnetostriction and magnetization at saturation, respectively. The behavior of $\lambda/\lambda_S$ in Fig 3(c) is compatible with magnetostriction measurements reported for directionally casted $Fe_{72.5}Ga_{27.5}$ [27].

In order to investigate the reproducibility of the switching behavior, we performed a "training" of the contact by employing a large number of open-close cycles shown in Fig. 4. $n_0$ corresponds to several hundred cycles performed during several hours. Apart from the low-field conductance switching mentioned above, we observe a decrease of the switching field and of the hysteresis with increasing cycle number $n > n_0$, see Fig. 4 (inset). The linear decrease of the switching field observed in both directions, i.e., for opening and closing, indicates a modification of the contact configuration by the "training effect". In fact, the switching field necessary for closing the contact decreases slightly faster with increasing $n$ than the switching field for opening the contact. The hysteresis $\Delta H$ between both switching fields decreases continuously from 1.4 mT to 1.2 mT after 6 further cycles. These values are considerably smaller than for previously reported magnetostrictive MCBJs made from dysprosium [13] ($\mu_0\Delta H \approx 20$ mT) and Terfenol-D [14] ($\mu_0\Delta H \approx 100$ mT).

In addition to the magnetic-field induced switching of the contact, the mechanical switching by using the piezo-driven pushing rod was also studied. Fig. 5 shows the variation of the conductance while closing the contact for up to 186 cycles (only a subset of data is shown for clarity). Starting from an open contact, the conductance gradually increases for $G/G_0 << 1$ which is attributed to electron tunneling between the two electrodes, see below. Further reduction of the electrode gap leads to a "jump to contact", see curves for $n \geq 72$ in Fig. 5 and Fig. 6, after which the conductance increases further and shows a number of steps and conductance plateaus. The latter are characteristic for few-atom contacts and indicate a change of the atomic configuration. The deviation of $G$ on the plateaus from integral multiples of the conductance quantum is attributed to electron backscattering at defect sites near the contacts [7]. The inset of Fig. 5 shows the histogram of conductance values[7] (0.1 bin size) obtained from a large number of conductance curves. The histogram shows a broad maximum centered at $G/G_0 \approx 0.8$ for opening the contact. In this case, a single-atom contact is usually established just before the contact breaks when increasing the electrode distance. This maximum is missing when closing the contact, for which the histogram is dominated by the large number of contacts with $G/G_0 << 1$ due to electron tunneling discussed below. This may be due to the fact that the closing of the junction occurs more gradually with the slow increase of tunneling current as the contacts are brought close together [13].

The detailed behavior in the tunneling regime observed for $G(V)$ and $G(H)$ while closing the contact is shown in Fig. 6. In both cases, the conductance increases with decreasing electrode separation $\Delta x$, i.e., with decreasing piezo voltage ($V \sim \Delta x$) or increasing field/magnetization. Fig. 6 (inset) shows semi-logarithmic plots of the conductance $G/G_0$ vs. piezo voltage and $G/G_0$ vs. magnetic field obtained on the same sample. The linear behavior of ln ($G/G_0$) in both cases is characteristic for electron tunneling [28]. The relation between electrode distance $\Delta x$ and piezo voltage can be obtained from the slope of ln $[G(V)/G_0]$ using $G \sim \exp(-\Delta x/\xi_V)$, where $\xi_V = 0.36$ Å takes into account the average work function of the electrode in helium (7.55 eV) [14, 29].

By the same token, the effect of the magnetic field on the conductance in the tunneling regime is described by $G \sim \exp(-\mu_0 H/\xi_H)$, where $\xi_H$ is determined from the linear slope of ln $[G(H)/G_0]$ shown in Fig. 6 (inset). From the data plotted in Fig. 6 we obtain $\xi_H = -0.086$ mT and, hence, $\Delta x/\Delta(\mu_0 H) = \xi_V/\xi_H = -420$ nm/T. The negative sign is in agreement with the fact that the gap decreases by the elongation of each electrode due to the increasing magnetostrictive strain in increasing field. $\Delta x/\Delta(\mu_0 H)$ is similar to the value obtained for Dy nanocontacts [13] but two orders of magnitude larger than for $Tb_{0.3}Dy_{0.7}Fe_{1.95}$ (Terfenol-D) nanocontacts [14]. The difference is attributed to the strong anisotropy of the magnetostrictive strain and the high strength of the applied field in the case of Terfenol-D. Essentially, $\Delta x/\Delta(\mu_0 H)$ is roughly proportional to the variation of the magnetostrictive strain with field $d\lambda/dH$. For the present case, as well as for Dy [13], the magnetic field was much smaller than the saturation field (large $d\lambda/dH$) and was applied along the wire or ribbon axis, respectively. In contrast, for Terfenol the tunneling measurements were done at a high field of $\sim 1$ T almost at saturation (small $d\lambda/dH$) with the field applied perpendicularly to the wire axis [14]. Eventually, the saturation magnetostriction $\lambda_s$ can be estimated by taking into account the qualitative behavior of $\lambda(M)$ [Fig. 2(b)] [14]. For $Fe_{73}Ga_{27}$ nanocontacts we obtain $\lambda = 4 \times 10^{-5}$, which is ten times smaller than observed for $Fe_{81}Ga_{19}$ single crystals along the <100> direction at $T = 4$ K [30]. However, in $Fe_{1-x}Ga_x$ polycrystals the average magnetostriction is presumably strongly reduced due to the different orientation of grains and the contribution from the negative magnetostriction along the <111> direction [30]. For the investigated $Fe_{73}Ga_{27}$ samples, the grains are preferentially oriented with the <110> direction along the ribbon axis which is between <100> and <111>.

**Conclusion**

In summary, we have demonstrated the operation of a magnetostrictive nanocontact device made from low-field magnetostrictive $Fe_{73}Ga_{27}$ (Galfenol) which allows to reproducibly switch the conductance between high and zero conductance in a low magnetic field at 10 K. The results represent a mark in the field of magnetic devices that rely on magnetostrictive materials in future nanotechnology applications. The very low hysteresis of ~ 1 mT for conductance switching between opening and closing the contact is in accordance with the $M(H)^2/M_s^2$ behavior. Tuning the switching field, i.e., the field where the contact switches between `on´ and `off´ states, was demonstrated by performing a training of the contact which gives rise to an optimized contact geometry with respect to low loss conductance switching. Investigation of the tunneling behavior of a mechanically and magnetically operated contact at low temperature allows an estimation of the low-temperature magnetostriction $\lambda = 4 \times 10^{-5}$ of this material.

**Acknowledgements**

SNJ would like to thank the Deutscher Akademischer Austauschdienst (DAAD) – Indian Institute of Technology (IIT) Faculty exchange program and IIT Hyderabad for financial assistance. SNJ would also like to thank Department of Science and Technology (#SR/FTP/PS-190/2012), India for funding.

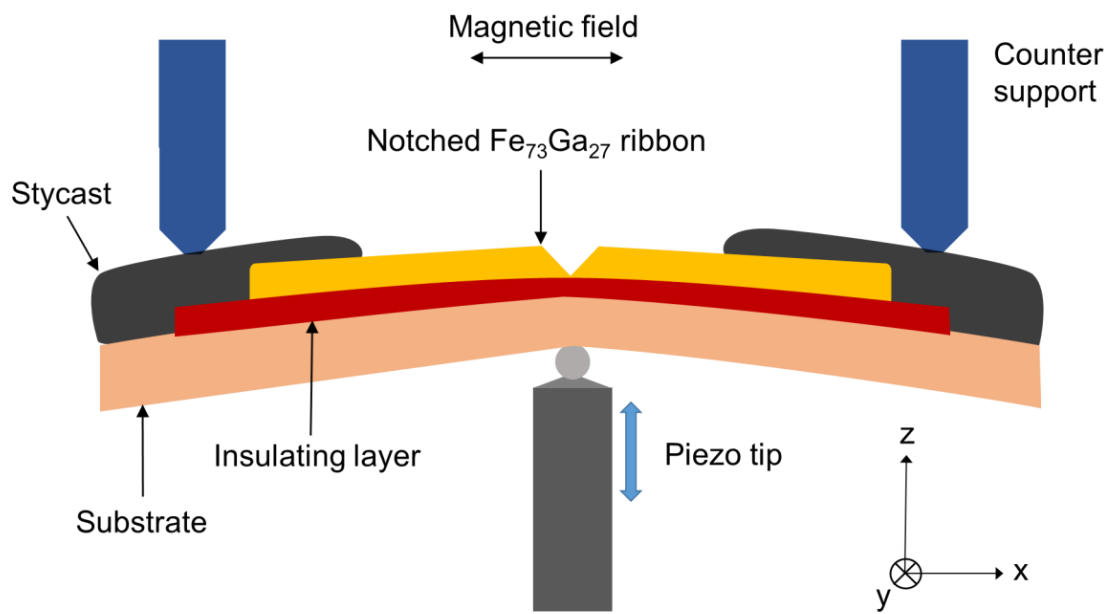

**Fig. 1:** Schematic diagram of a mechanically controlled break-junction (MCBJ) device. A Fe$_{73}$Ga$_{27}$ ribbon sample (yellow) is glued to a flexible Cu-bronze substrate (orange) using Stycast epoxy (black). An insulating layer (red) prevents direct electrical contact between the ribbon and the substrate. The substrate and the ribbon can be reversibly bent using a voltage controlled piezo tip (grey) which pushes the substrate against two counter supports (blue).

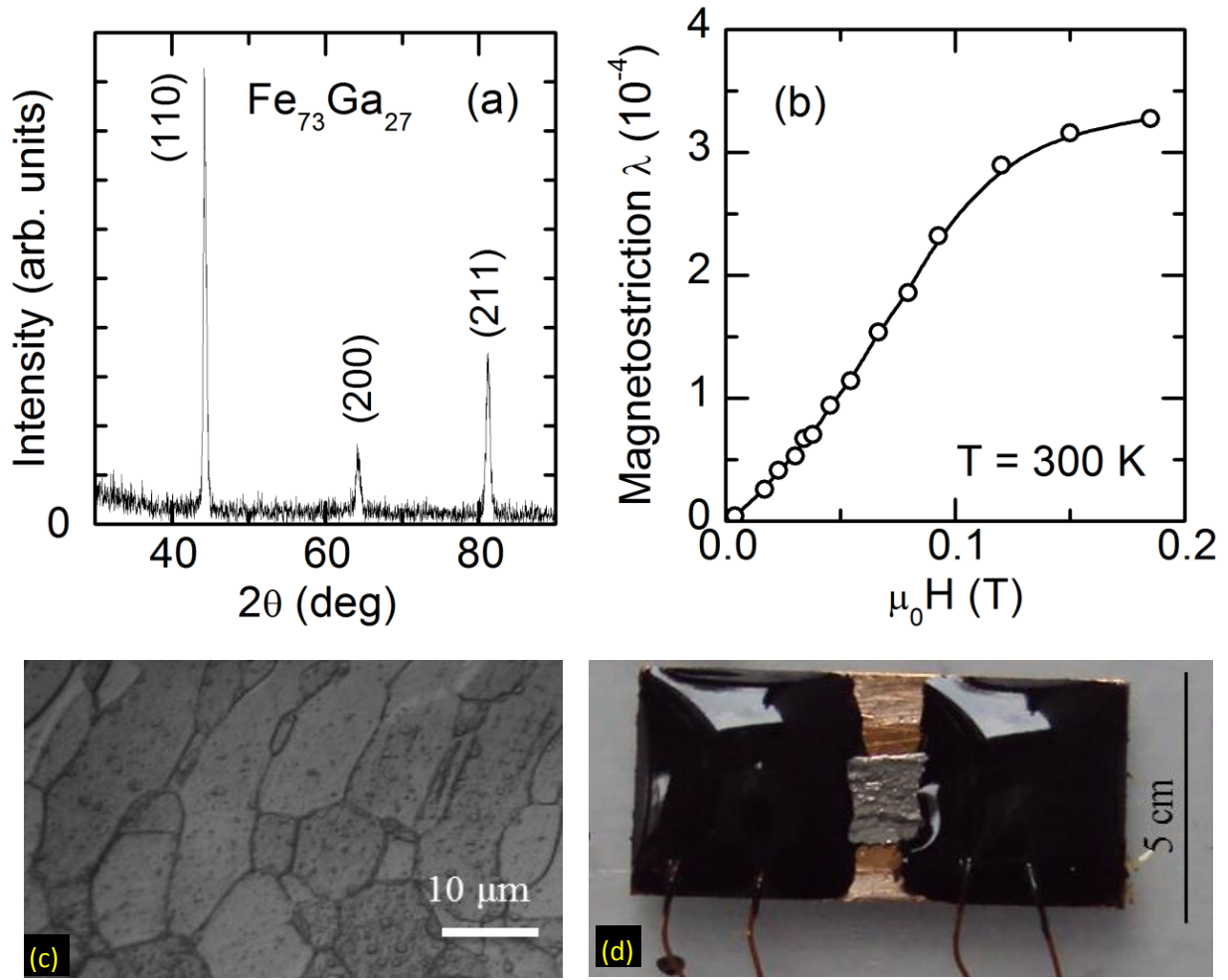

**Fig. 2:** (a) X–ray diffraction pattern of a $Fe_{73}Ga_{27}$ ribbon. Miller indices of the Bragg reflections refer to the A2 phase of bcc Fe. (b) Magnetostriction $\lambda(H)$ at $T = 300$ K. (c) SEM image of a ribbon. (d) Photograph of the MCBJ device.

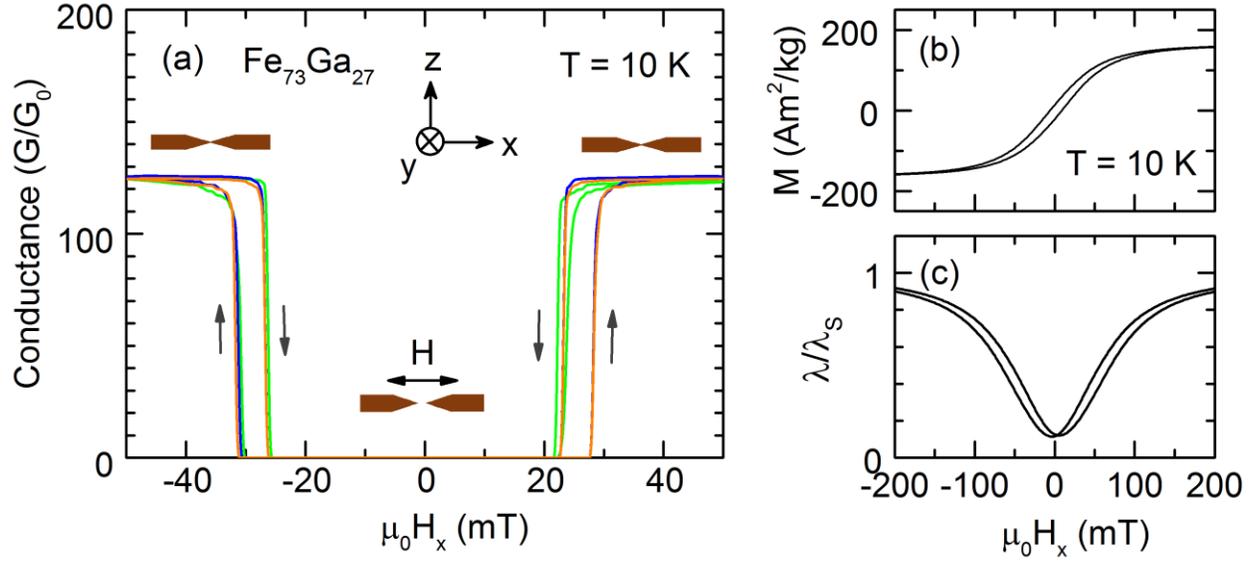

**Fig. 3:** (a) Conductance $G/G_0$ vs. magnetic field $H_x$ of a MCBJ prepared from a Fe$_{73}$Ga$_{27}$ ribbon. The magnetic field is applied along the $x$ direction of the ribbon. Only 3 cycles out of 13 are shown for clarity. Cartoons indicate the open and closed configurations of the nanocontact. (b) Magnetization $M(H_x)$. (c) Magnetostriction $\lambda/\lambda_s = (M/M_s)^2$, derived from $M(H_x)$, along the x direction.

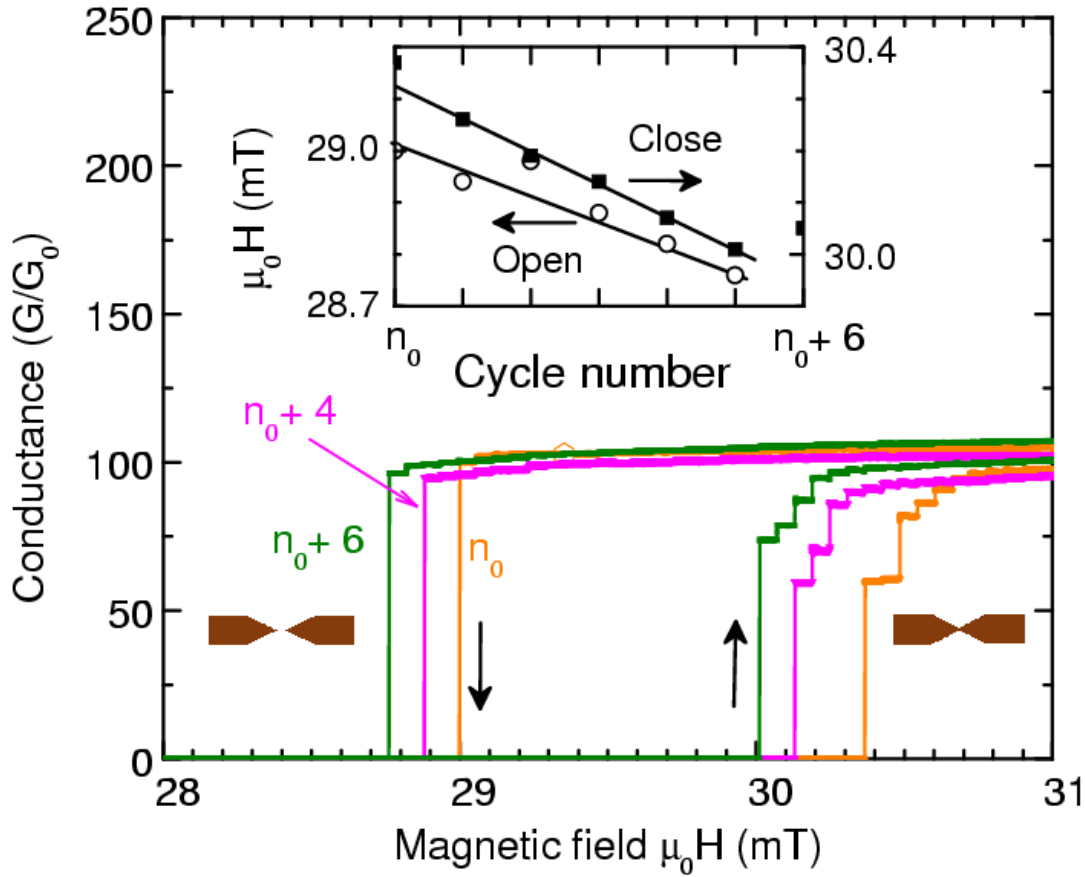

**Fig. 4:** Conductance $G(H)/G_0$ of a $Fe_{73}Ga_{27}$ ribbon for consecutive open-close switching cycles. Inset shows the variation of switching field vs. cycle number for closing and opening of the contact. The switching field decreases almost linearly with increasing cycle number. Cartoons show the open and closed configurations of the nanocontact.

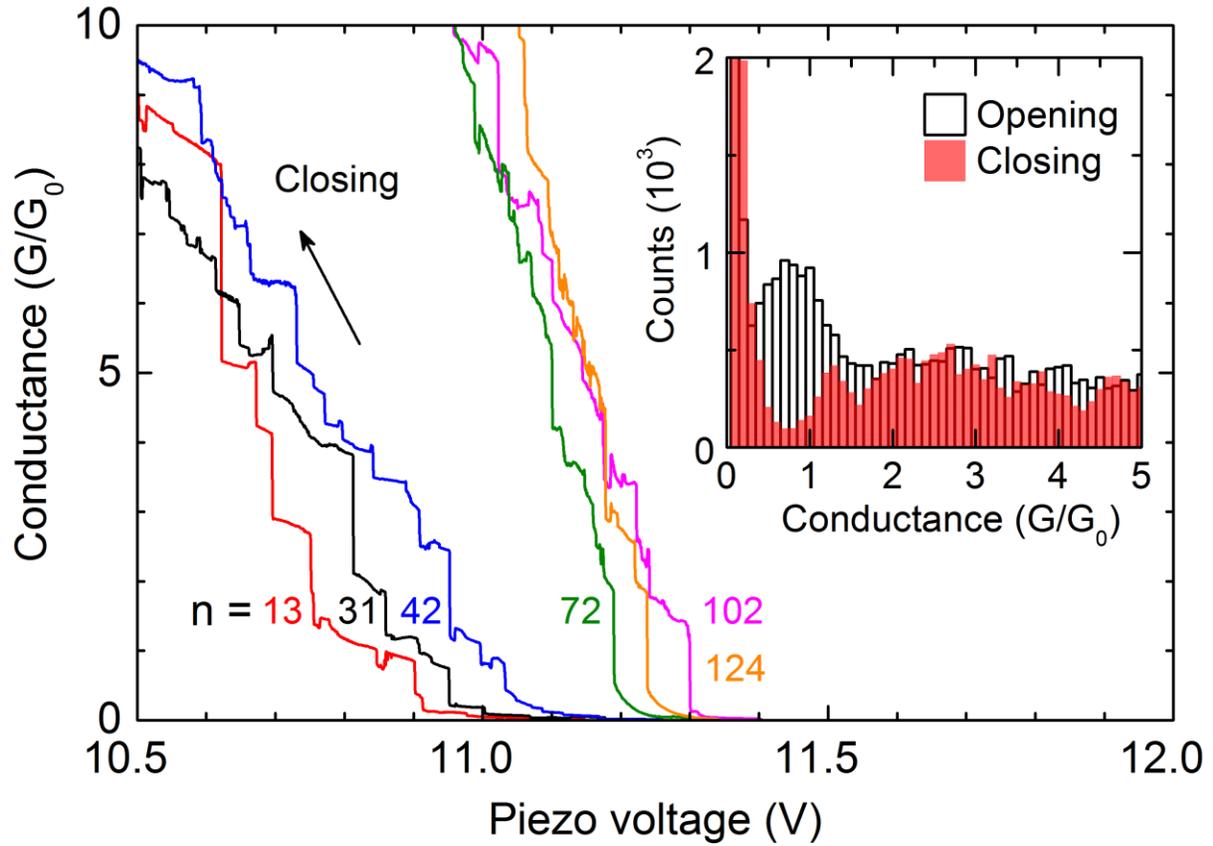

**Fig. 5:** Conductance ($G/G_0$) vs. piezo voltage, i.e., electrode separation, while closing the junction mechanically for different cycles $n$. Inset shows the distribution of counts for various $G$ values during opening (white) and closing (red) of the junction.

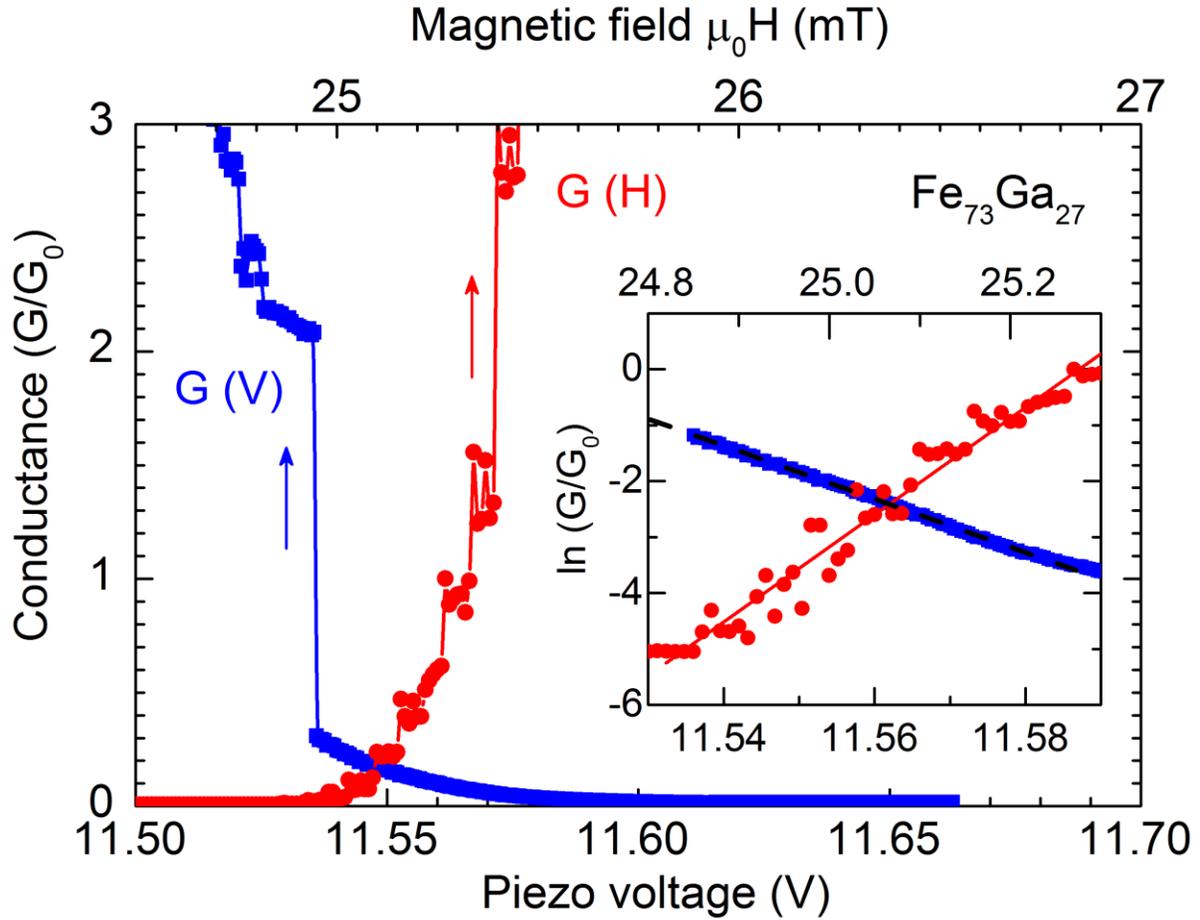

**Fig. 6:** Conductance $G/G_0$ vs. piezo voltage (blue) and vs. magnetic field $H$ (red) during closing a nanocontact which exhibits electron tunneling for $G/G_0 \ll 1$ and a "jump to contact" at piezo voltage ~ 11.54 V or at 25.3 mT. Inset shows semi-logarithmic plots of the data. Broken and solid lines indicate a behavior $G \sim \exp(-\Delta x/\xi_V)$ or $G \sim \exp(-\Delta H/\xi_H)$, respectively with $\xi_V > 0$ and $\xi_H < 0$.